\documentclass[journal=jacsat,manuscript=article]{achemso}
\usepackage[version=3]{mhchem} % Formula subscripts using \ce{}

\author{Halimah Harfah}
\altaffiliation{Both authors contributed equally to this work}
\affiliation[Osaka University]
{Graduate School of Engineering Science, Osaka University, 1-3 Machikaneyama-cho, Toyonaka, Osaka 560-8531, Japan}
\alsoaffiliation[Universitas Indonesia]
{Department of Physics, Faculty of Mathematics and Natural Science, Universitas Indonesia, Kampus UI Depok, Depok, Jawa Barat 16424, Indonesia}
\author{Yusuf Wicaksono}
\altaffiliation{Both authors contributed equally to this work}
\affiliation[Osaka University]
{Graduate School of Engineering Science, Osaka University, 1-3 Machikaneyama-cho, Toyonaka, Osaka 560-8531, Japan}
\author{Muhammad A. Majidi}
\affiliation[Universitas Indonesia]
{Department of Physics, Faculty of Mathematics and Natural Science, Universitas Indonesia, Kampus UI Depok, Depok, Jawa Barat 16424, Indonesia}
\author{Koichi Kusakabe}
\email{kabe@mp.es.osaka-u.ac.jp}
\phone{+81 (0)6-6850-6406}
\fax{+81 (0)6-6850-6406}
\affiliation[Osaka University]
{Graduate School of Engineering Science, Osaka University, 1-3 Machikaneyama-cho, Toyonaka, Osaka 560-8531, Japan}

\title{Spin-Current Control by Induced Electric-Polarization Reversal in Ni/hBN/Ni: A Cross-Correlation Material}


\begin{document}

\begin{abstract}
We undertook an ab-initio study of hexagonal boron nitride (hBN) sandwiched between Ni(111) layers to examine the interface of this material structure. We considered Ni(111)/hBN/Ni(111) with a slab with three Ni atomic layers to determine the exact atom arrangement at the interface. The density functional theory calculations for 36 stacking arrangements, which are doubled with respect to the magnetic alignment of slabs in an anti-parallel configuration (APC) and parallel configuration (PC), revealed that the number of formed weak chemical bonds, in the pd-hybridization between the N and Ni atoms, is decisive. A maximum of two pd-hybridization bonds stabilized the structure, with APC proving to be the most favorable magnetic alignment, in line with the results of previous experimental studies. In the lowest energy state, an induced magnetic moment at an N site appears when N is moved closer to one of the Ni atoms. Interestingly, the moment direction is switched by the position of the N layer in the resulting bi-stable state with electrical polarization when APC is chosen. The transmission probability calculation of Ni/hBN/Ni having the determined interface structure at the center of the junction exhibits a spin-filtering effect where the spin-polarized current is controlled by the electric field when a field-induced reversal of the polarization is realized.
\end{abstract}

\section{Introduction}
The magnetic tunnel junction, the so-called "spin-valve structure," is vital to the development of spintronic devices capable of fulfilling the high demand for applications based on a spin configuration. Despite many studies of the spin-valve structure having been undertaken \cite{Dankert:2015,Martin:2015,Iqbal:2015,Wang:2015,Wu:2015,Banci:2016,Caneva:2015,Caneva:2016,Iqbal:2016,Asshoff:2017,Entani:2016,Iqbal:2016-JMCC,Cobas2016}, indicating the potential of the devices' application to logic devices, hard-drive magnetic read heads, and magnetic sensors  \cite{Chappert:2007,Dery:2007,Zhu:2006,Childress:2005,Chen:2010,Iqbal:2018-review}, considerable improvement is still required to attain a higher level of magnetoresistance. Fully understanding the nature of the properties of this device is important to enable successful design and application. 
Owing to the striking characteristics of two-dimensional (2D) materials, graphene and related 2D materials have been examined for use as non-magnetic spacers in current-perpendicular-to-plane spin-valve devices \cite{Banci:2016,Wang:2015,Iqbal:2018-review}. 
By using 2D materials for the non-magnetic spacer, the problem of controlling the flatness of the surface at the interface when improving and downscaling devices can be tackled, because graphene offers perfect flatness while being only one atomic layer thick. 
A theoretical study revealed that graphene can be used as a spin filter \cite{Karpan:2007}. Several experimental studies addressing the fabrication of graphene-based spin-valve devices through the application of different fabrication methods such as exfoliation \cite{Exfolation1,Asshoff:2017}, wet thermal techniques\cite{Iqbal:2015-JMCC,Iqbal:2013-NanoRes,Chen:2013,Li:2014,LiWan:2014}, hydrothermal techniques\cite{Mandal:2012}, and direct CVD \cite{Martin:2015,Entani:2016} have been reported.
Although the magnetoresistance effect was observed, the results were still unsatisfactory. This could be a result of the high conductivity and lack of bandgap of graphene, thereby increasing the demand for 2D-based semiconducting or insulating materials.

Hexagonal boron nitride (hBN), a graphene-like structure with a large bandgap and weak spin-orbit coupling, can be a perfect candidate for this spin-valve junction. Experiments comparing the magnetoresistance performance of graphene and hBN were undertaken using thin Ni films for the magnetic layers, sandwiching either graphene or hBN \cite{Iqbal:2018-Ni-hBN-Ni_exp}.
Among other transition metal surfaces, Ni(111) was chosen because it has the smallest lattice mismatch with both graphene and hBN. By using Ni as the ferromagnet and hBN as the interlayer, a perfect interface can be achieved. The magnetoresistance measurement results show that the magnetoresistance of Ni/hBN/Ni is higher than that of the Ni/graphene/Ni spin-valve \cite{Iqbal:2018-Ni-hBN-Ni_exp}. 
An extensive and intensive theoretical investigation of the Ni/graphene/Ni spin-valve structure was undertaken, leading to the possibility of developing a new spintronic device design \cite{Wicaksono:2019}. Since relatively few theoretical studies of Ni/hBN/Ni have been undertaken, a full understanding of the material properties is yet to be attained.

This paper presents the first comprehensive investigation of different configurations of a Ni/hBN/Ni spin-valve structure: a stacking arrangement and its impact on the magnetic response and transport properties of Ni/hBN/Ni nanostructures. All 36 stacking arrangements, twice the number with respect to the magnetic alignment of the slabs in an anti-parallel configuration (APC) and parallel configuration (PC), are considered. The stability of the stacking, as well as the electronic and magnetic properties, is determined using generalized gradient approximations of density functional theory. After the geometry relaxation, each system specified by a stacking arrangement reaches the equilibrium state.

All determined structures are found to exhibit a unique tendency, according to which the structures are grouped into three, based mainly on the total energy range. These groups are the lowest total energy or the most stable state (group I), a total energy of $\approx$ 100 meV (group II), and  $\approx$ 400 meV (group III) relative to the most stable state. The distance between the B-N atom and the nearest Ni atom shows that the existence of the group can be associated with the bond formation resulting from the pd-hybridization between the N and Ni atoms. Among the three groups, the first group has two stacking arrangements, namely asymmetric and symmetric. Therefore, in the present study, the calculation of the magnetic response and transport properties is undertaken for four structures that represent each group (two for group I and one for each of the other groups). By performing spin-charge density mapping, we find a bi-stable state in the asymmetric stacking arrangement (group I). This type of stacking arrangement leads to a structural deformation from a flat hBN plane to a rugged hBN plane. The buckling direction is two-fold and can be tuned by applying an external electric field. When the buckling direction is switched, the induced dipole moment in the BN layer is also switched to have a reversed dipole. Thus, these results reveal the uniquely different and unprecedented characteristics of Ni/hBN/Ni, providing new insights for prospective materials based on spin-current control by induced electric-polarization reversal. 
Meanwhile, the transmission probability results show the effects of the spin-valve on groups I and III.
The results reveal that the former has a higher magnetoresistance ratio than the latter. Group III, however, can switch between the APC and PC states more easily than group I. Therefore, the different characteristics of groups I and III may lead to the consideration of two types of applications: cross-correlation devices and magnetic sensors.

\section{Results and discussion}

\subsection{Total energy and groups of stacking arrangements}

\begin{figure}[tb]
\centering
\includegraphics[width=\linewidth]{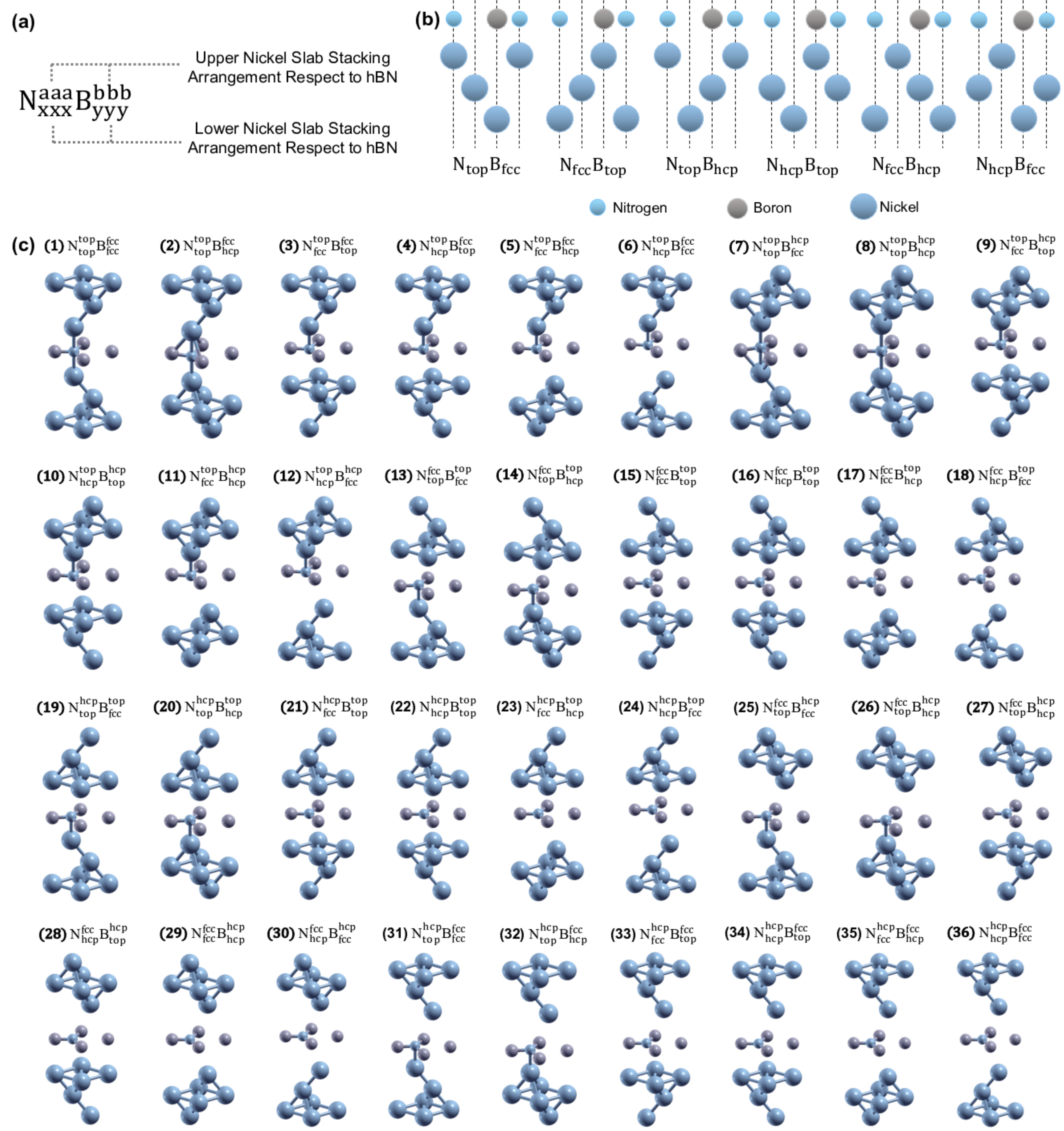}
\caption{Thirty-six stacking arrangements with stacking arrangement name and configuration number.}
\label{fig:1}
\end{figure}

\begin{figure}[tb]
\centering
\includegraphics[width=\linewidth]{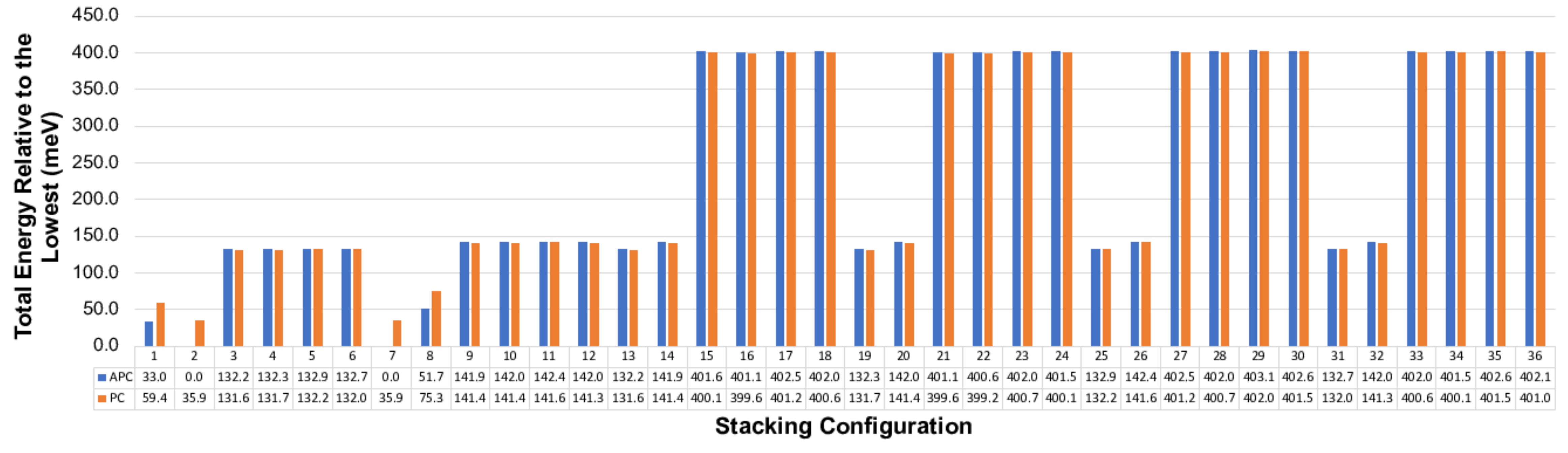}
\caption{Histogram and table of total energy for each stacking arrangement in APC and PC.}
\label{fig:2}
\end{figure}

The spin-valve is constructed by sandwiching a single layer of hBN with three layers of Ni. The $<0.4\%$ lattice mismatch between the hBN and Ni layers assumes that both are commensurate. For the Ni-hBN interface, it is already known that there are six possible arrangements of Ni and B-N atoms. 
This arrangement is based on three different symmetry sites: top (either B or N placed on top of the first Ni layer), fcc (either B or N placed on top of the third Ni layer), and hcp (either B or N placed on top of the second Ni layer), as shown in figure \ref{fig:1}.(b). Here, the N$^{\textrm{a}}_{\textrm{x}}$B$^{\textrm{b}}_{\textrm{y}}$ symbol indicates that the N and B atoms are placed in the "a" and "b" sites of the upper nickel slab, respectively, and in the "x" and "y" sites of the lower nickel slab. The site names of "a," "b," "x," and "y" are top, fcc, or hcp. Because of the two interfaces in the spin-valve, there are 36 possible arrangements in total (figure \ref{fig:1}). 
Figure \ref{fig:2} shows the total energy relative to the lowest of the 36 stacking arrangements in the APC and PC states. The results also show that, according to the total energy of the PC and APC, the arrangements can be classified into three groups.
The first group (group I) is a stacking arrangement with a total energy of 0-60 meV, relative to the lowest energy state (also regarded as being the most stable state). The stacking arrangements of configurations 2 and 7 show the lowest total energy, which indicates that the N$^{\textrm{top}}_{\textrm{top}}$B$^{\textrm{fcc}}_{\textrm{hcp}}$ and N$^{\textrm{top}}_{\textrm{top}}$B$^{\textrm{hcp}}_{\textrm{fcc}}$ stacking arrangements correspond to the most stable state. These two stacking arrangements have an asymmetric configuration of Ni(111) slabs, above and below the hBN layer. While configurations 1 and 8, which arise from the symmetric arrangement of N$^{\textrm{top}}_{\textrm{top}}$B$^{\textrm{fcc}}_{\textrm{fcc}}$, while N$^{\textrm{top}}_{\textrm{top}}$B$^{\textrm{hcp}}_{\textrm{hcp}}$ has a slightly higher total energy with a difference of $\approx$ 30 meV. The small total energies of these configurations arise from the two strong pd-hybridizations between the p$_z$-orbital of the N atoms and d$_{z^2}$-orbital of both the upper and lower Ni atoms at the interfaces, which has also been reported for the Ni(111)/hBN interface \cite{Joshi:2013-hBN-Ni_theory}. 

This hybridization can be observed in the interlayer distance between the upper or lower Ni(111) slab and the hBN layer. A previous experimental and theoretical study of the Ni(111)/hBN interface suggested that the interlayer distance between the Ni(111) slab and the hBN layer is $1.87 \pm 0.12 $ \AA \cite{Tonkikh:2016-hBN-Ni_Interlayer_distance,Joshi:2013-hBN-Ni_theory}. In these configurations, the interlayer distance between either the upper or lower Ni(111) slab and the hBN layer is $\approx 2.10$ \AA~, as shown in table \ref{table:1}, indicating the relevant distance of the bonding nature. The small discrepancy between the experimental result obtained for the Ni(111)/hBN interface and the theoretical result obtained for Ni/hBN/Ni may originate from the different number of pd-hybridizations. On the Ni/hBN interface, pd-hybridization is formed between a N atom and one Ni atom, while for the Ni/hBN/Ni of group I, the N atom forms pd-hybridizations with two Ni atoms from the upper and lower Ni(111) slabs, leading to a reduction in the bond strength.  
Moreover, although both the asymmetric and symmetric configurations have two pd-hybridizations, their total energies are slightly different. This difference originates in the strength of the covalent bond due to pd-hybridization, where the asymmetric configuration has a stronger bond than the symmetric one, as can be determined from the shorter interlayer distance. Interestingly, within the asymmetric configuration itself, the APC and PC states have different bonding formations. In the APC state, a weak covalent bond is formed between the Ni and B atoms at the interface. This bond formation breaks the flatness of the hBN layer and yields a rugged plane with the distance between the B and N atoms in the z-direction being $0.25$ \AA

\begin{table}[ht]
\centering
\begin{tabular}{|l|l|l|l|l|l|l|l|l|l|}
\hline
\textbf{Layer} & \multicolumn{8}{l|}{\textbf{Interlayer distance (\AA)}} \\
\cline{2-9}
& \multicolumn{2}{l|}{\textbf{2pd asymmetry}} &  \multicolumn{2}{l|}{\textbf{2pd symmetry}} & \multicolumn{2}{l|}{\textbf{1pd}} & \multicolumn{2}{l|}{\textbf{no hybridization}} \\
\cline{2-9}
&\textbf{APC} & \textbf{PC} & \textbf{APC} & \textbf{PC} &\textbf{APC} & \textbf{PC} & \textbf{APC} & \textbf{PC} \\
\hline
N-Upper Ni slab & 2.13 & 2.14 & 2.16 & 2.19 & 2.07 & 2.07 & 3.30 & 3.30 \\
\hline
N-Lower Ni slab & 2.12 & 2.17 & 2.16 & 2.19 & 3.30 & 3.30 & 3.30 & 3.30 \\
\hline
\end{tabular}
\caption{Interlayer distance between upper and lower Ni(111) slabs with hBN layer for the representative of each stacking arrangement group in APC and PC.}
\label{table:1}
\end{table}

The structural deformation from a flat hBN plane to a rugged hBN plane structure can occur as a result of the charge distribution characteristics. This can be due to the bond formation leading to a different charge-transfer mechanism.  These bond formations among three B atoms, one N atom, and the Ni atom result in a high coordination of Ni. The Ni-N bond length in this asymmetric structure is $2.12$ \AA, which is the shortest among the structures having 2pd-hybridizations (group I). Therefore, for group I, the charge transfer from Ni to N is larger for the asymmetric structure than that for the symmetric structure. Assume that a B$_3$N structure is isolated from the bulk. Then, when the transferred charge goes into a dangling bond to create a lone pair of a N atom with three-fold coordination, the structure acquires sp$^3$ bonding. A typical example of such a mechanism is NH$_3$. The structure of NH$_3$ is not planar but rather a trigonal pyramidal shape. Occupation by a larger number of electrons gives rise to this shape. Our hBN plane, sandwiched between the Ni layers, exhibits a similar pyramidal structure, although this is caused by a larger electron occupation than that of a flat one, which is consistent with the fact that the Ni-N distance is smaller in the case of the pyramidal structure. The pyramid has two-fold direction. Since a negatively charged N atom and a positively charged B atom create a dipole moment in the axis of the pyramid along the Ni-N direction, a reversed dipole is formed when the pyramidal shape is inverted.

Moreover, the other energy ranges are of a different order compared to the lowest one. Configurations 3-6, 9-14, 19, 20, 25, 26, 31, and 32, which are in an order of $\approx$ 100 meV, are classified as being in the less-stable state (group II) with at least one pd-hybridization occurring between the p$_z$-orbital of the N atoms and either the upper or lower d$_{z^2}$-orbital of the Ni atoms at the interface. A single pd-hybridization can be inferred from the interlayer distance between the upper and lower Ni(111) slabs and the hBN layer. Table \ref{table:1} shows that the interlayer distance for the Ni(111) slabs with stacking arrangement N$_{\textrm{top}}$(N$^{\textrm{top}}$) is around $\approx 2.07$ \AA~, which is in agreement with the results of a previous study on the hBN/Ni(111) interface (which only has one pd-hybridization). By contrast, the other Ni(111) slab, for which the stacking arrangement is N$^{\textrm{hcp/fcc}}$(N$_{\textrm{hcp/fcc}}$), has an interlayer distance of $\approx 3.20$ \AA~, representing the van der Walls bonding nature. The last group (group III), that is, configurations 15-18, 21-24, 27-30, and 33-36, is the least stable structure of all groups with a stacking arrangement. Group III with a total energy range of $\approx$ 400 meV exhibits no pd-hybridization regardless of the configuration. The interlayer distance between the upper (lower) Ni(111) slab and the hBN layer is $\approx 3.28$ \AA. The absence of pd-hybridization arises from the fact that no Ni atom is placed either directly on top or below a N atom. However, interestingly, the total energies of N$_{\textrm{hcp/fcc}}^{\textrm{hcp/fcc}}$B$_{\textrm{top}}^{\textrm{top}}$ and N$_{\textrm{hcp/fcc}}^{\textrm{hcp/fcc}}$B$_{\textrm{fcc/hcp}}^{\textrm{fcc/hcp}}$ in either a symmetric or asymmetric arrangement are in the same order. This means that the hybridization between the upper or lower Ni(111) slab and the hBN layer can only occur via the hybridization of the Ni atoms at the interface with the N atoms of hBN, implying that placing the first Ni layer immediately above or below a N atom is the key to increasing the stability of the structure.

\subsection{Optimum magnetic alignment and magnetic properties}

\begin{figure}[tb]
\centering
\includegraphics[width=\linewidth]{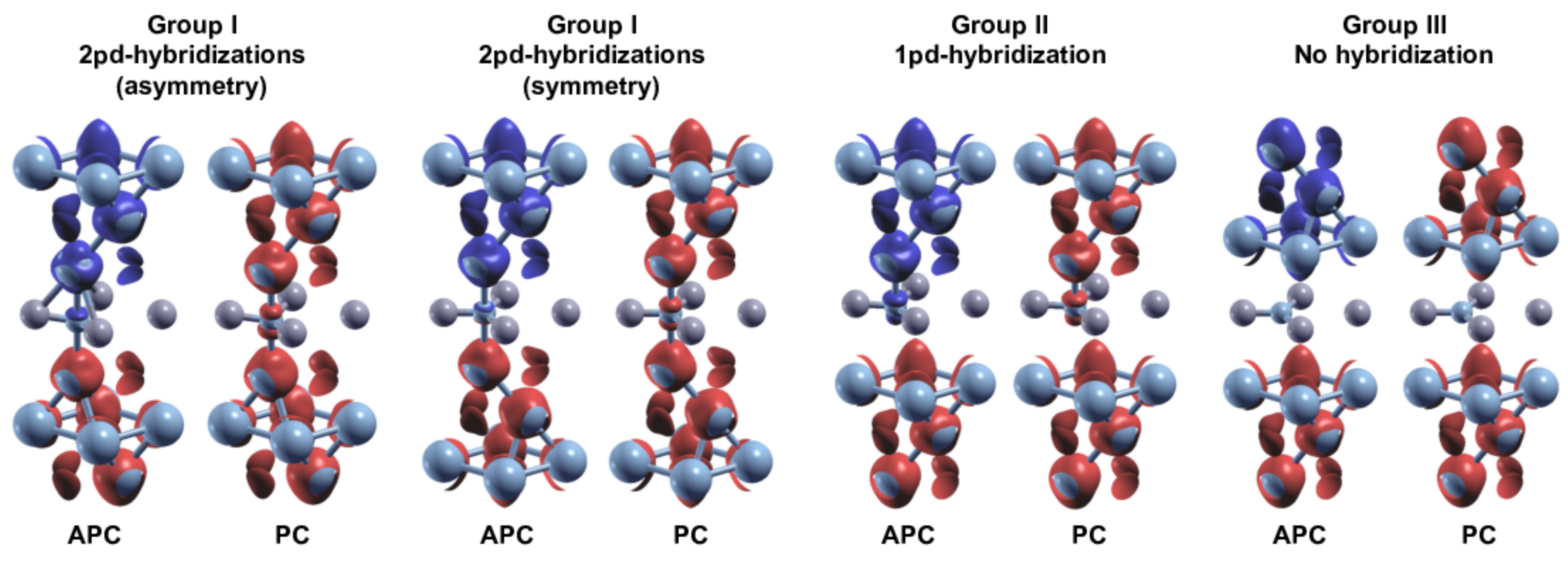}
\caption{Spin-charge density mapping for the representative of each stacking arrangement group having APC and PC states. The charge density in red represents spin-up electron density, while that in blue represents spin-down electron density.}
\label{fig:3}
\end{figure}

Figure \ref{fig:2} also shows that the APC state for the most stable structure (group I) has an energy level that is lower than that of the PC state. This result is in good agreement with those of previous experimental studies \cite{Iqbal:2018-Ni-hBN-Ni_exp}. This conclusion is further confirmed using different numbers of Ni layers, as described in table \ref{table:2}. This is in contrast with groups II and III, for which PC is more stable than APC. The total energy difference between these configurations also differs among groups I, II, and III, which have ranges of 23.7-35.9, 0.5-0.8, and 1.2-1.5 meV, respectively. The significant total energy difference for the most stable structure indicates that strong magnetic coupling occurs between the upper and lower Ni(111) slabs, mediated by the N atoms, and this strong coupling causes the system to be such that it is more difficult to flip the magnetic alignment from parallel to anti-parallel. The total energy difference for groups II and III is much smaller than that for group I, because there is less hybridization in the system.

\begin{table}[ht]
\centering
\begin{tabular}{|l|l|}
\hline
\textbf{Number of Nickel Layers} & {\textbf{E$_{\textrm{APC}}$ $-$ E$_{\textrm{PC}}$ (meV)}} \\
\hline
1 & 16.8 \\
2 & 22.3 \\
3 & 35.9 \\
4 & 21.9 \\
5 & 25.9 \\
6 & 25.3 \\
\hline
\end{tabular}
\caption{Total energies of different numbers of nickel layers having an asymmetric stacking arrangement (N$^{\textrm{top}}_{\textrm{top}}$B$^{\textrm{fcc}}_{\textrm{hcp}}$) relative to the ground state (APC).}
\label{table:2}
\end{table}

To further investigate the dependence of magnetic properties of Ni(111)/hBN/Ni(111) on the stacking arrangement, we determined the spin-charge density mapping for the representative of each stacking arrangement group. The results are shown in figure \ref{fig:3}. The spin-charge density mapping for group I shows that, in a parallel configuration, for either an asymmetric or a symmetric stacking arrangement, pd-hybridization leads to a charge transfer from the Ni atoms to the N atoms and results in an induced magnetic moment for the N atoms. However, in the APC state, the asymmetric and symmetric stacking arrangements exhibit different magnetic properties. The symmetric stacking arrangement shows that no induced magnetic moment is produced by the N atoms. This result arises from the nature of the exchange interaction between the Ni and N atoms, which prefer to have a parallel configuration \cite{Joshi:2013-hBN-Ni_theory}. Since the upper and lower Ni(111) slabs have opposite magnetic alignments, the induced magnetic moments from the upper and lower Ni(111) slabs cancel each other out, yielding zero magnetic moment. 

\begin{figure}[ht]
\centering
\includegraphics[width=\linewidth]{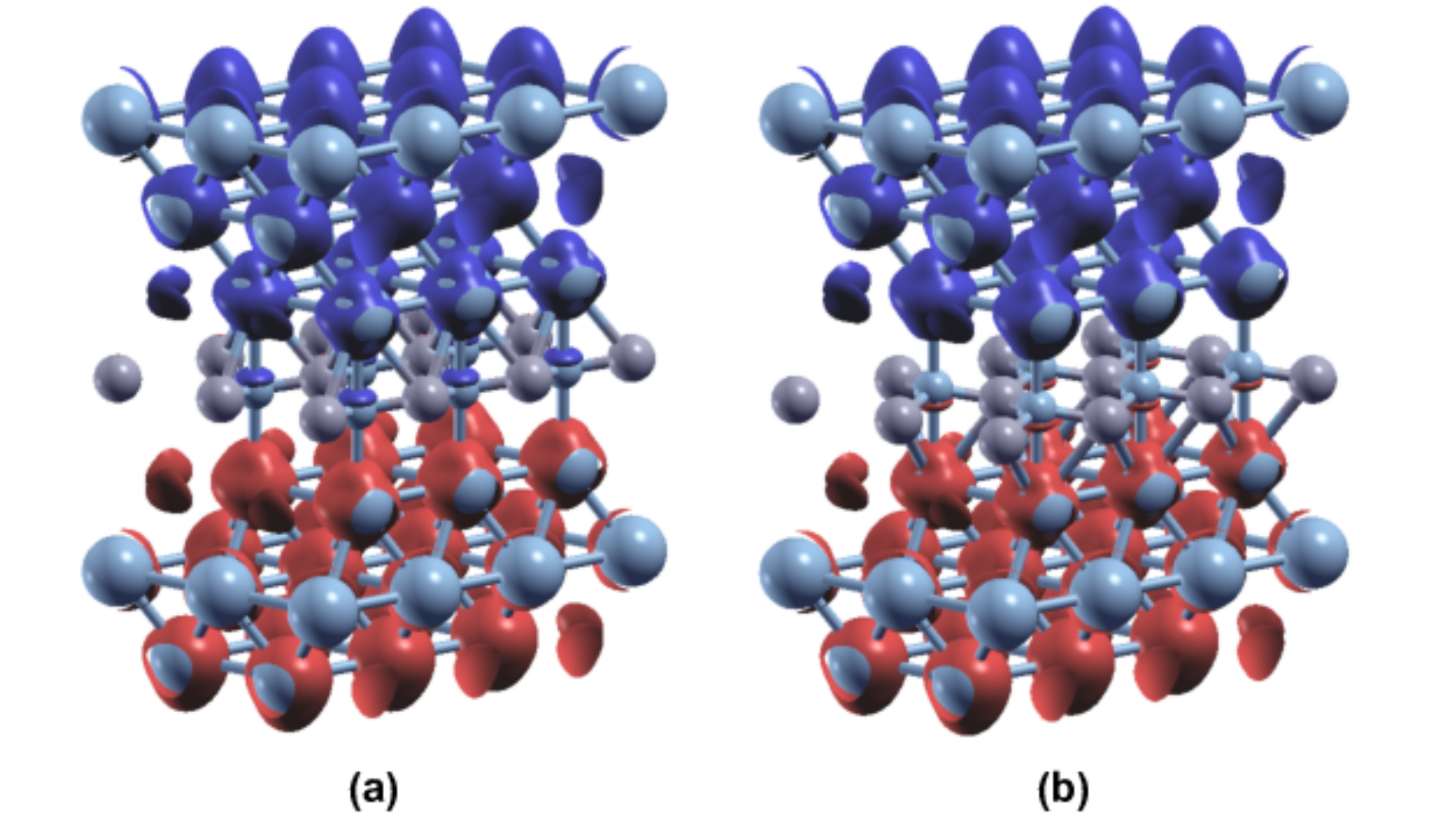}
\caption{Spin-charge density mapping of bi-stability state of the most stable stacking arrangement. The charge density in red represents the spin-up electron density, while that in blue represents the spin-down electron density.}
\label{fig:bistability}
\end{figure}

Meanwhile, the asymmetric stacking arrangement, which has a unique bi-stability state, exhibits a different induced magnetic moment response to N atoms in two different stable states. The direction of the induced magnetic moment occurs for N atoms parallel to the Ni(111) slab, with which B atoms have a weak covalent bond, as shown in figure\ref{fig:bistability}. Since the electric dipole moment occurs in the pyramid axis along the Ni-N direction, the induced magnetic moment for the N atoms can be controlled by applying an external electric field or electromagnetic field. This characteristic leads to a unique functionality, which will be discussed in greater detail later.

For the stacking arrangements of group II, the APC and PC states yield an induced magnetic moment that is aligned with the Ni(111) slab, with which the N atoms form a hybridization. This result is similar to that occurring at the Ni(111)/hBN interface \cite{Joshi:2013-hBN-Ni_theory}, since both systems have only one pd-hybridization. Meanwhile, for the stacking arrangement of group III, the induced magnetic moment does not occur for the hBN layer regardless of whether the system is in the APC or PC state. 

The spin-charge densities determined for all stacking arrangement groups indicate that the induced magnetic moment occurs in the hBN layer only when pd-hybridization is formed with the Ni(111) slab. The atoms that hybridize with the Ni(111) slab, i.e., N atoms, have an induced magnetic moment parallel to the Ni(111) slab. Meanwhile, for the 2-pd hybridizations (group I), the induced magnetic moment occurs when the stacking arrangement is asymmetric.

\subsection{Interface dependence of transmission probability}

\begin{figure}[ht]
\centering
\includegraphics[width=\linewidth]{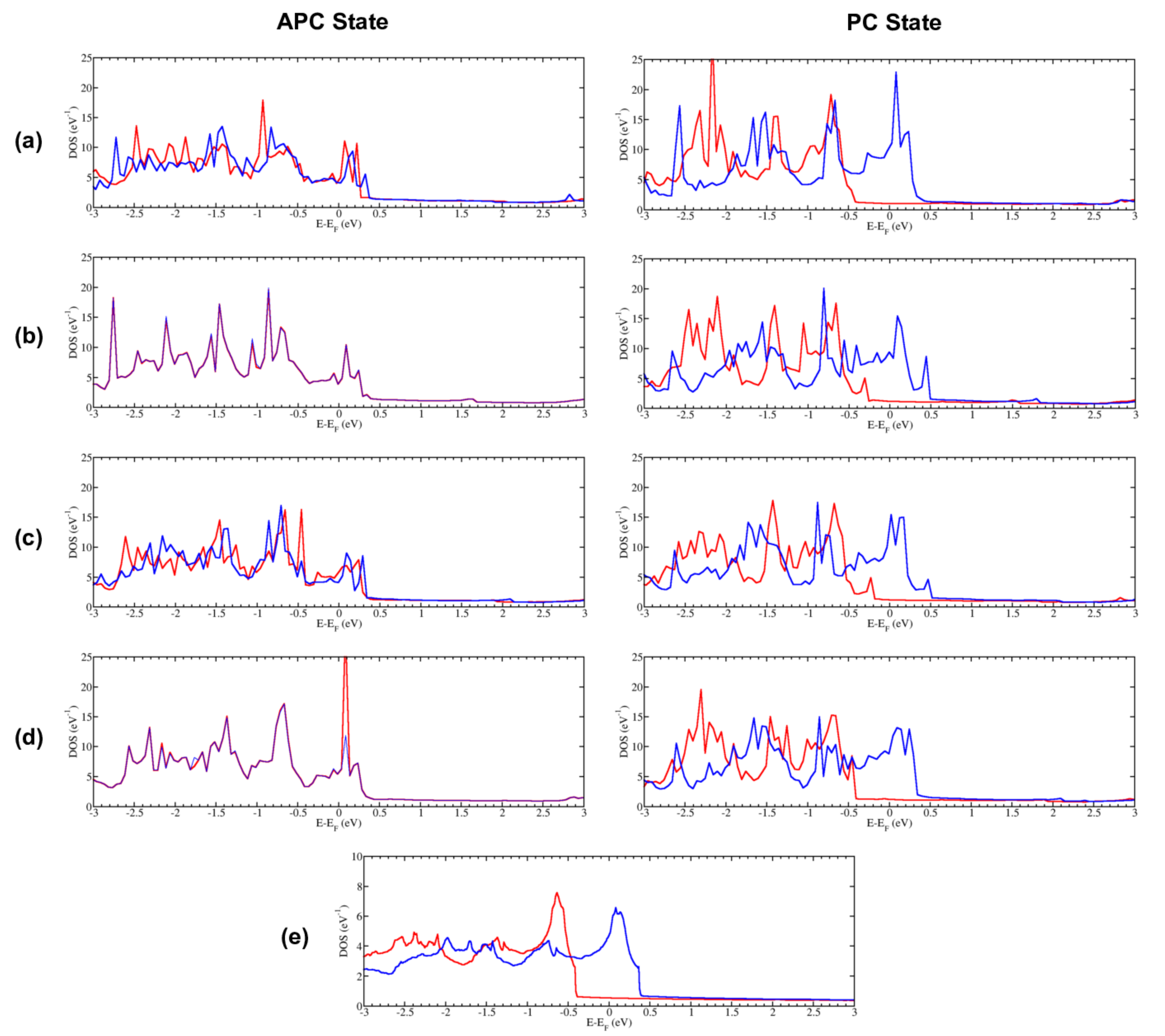}
\caption{Density of states (DOS) of 2pd-hybridizations: (a) asymmetric and (b) symmetric, (c) 1pd-hybridization, and (d) no hybridization stacking arrangement groups in APC and PC states for spin-up and spin-down electrons. The red line represents the spin-majority channel, while the blue line represents the spin-minority channel.}
\label{fig:4}
\end{figure}

The DOS of Ni(111) leads is shown in figure \ref{fig:4}.(e). From the DOS of the left or right lead, we can see that a Stoner gap appears, which is normal for a ferromagnetic material. The spin majority channel is shifted to a lower energy level, causing the spin minority to have a much higher density at the Fermi energy compared to the spin majority channel at zero temperature. Therefore, assuming that the Ni layers act as two electrodes in a magnetic tunnel device, the transmittance from the left to right leads is mainly a result of electrons in the minority spin channel. 

Meanwhile, the DOS values of Ni(111)/hBN/Ni(111) as the scatterer in different stacking arrangement groups are shown in figure \ref{fig:4}.(a), (b), (c), and (d). In the PC state, if we create a tunnel junction structure having Ni(111)/hBN/Ni(111) as a magnetic layer, all groups exhibit a spin-filtering effect for the spin-down electron when a non-spin-polarized current is injected through the junction. This is because the DOS at the Fermi energy is nearly zero for spin-up electrons and relatively high for spin-down electrons. Furthermore, in the APC state, the spin majority and minority channels overlap for groups I and III, which have a symmetric stacking arrangement with 2-pd hybridizations and no hybridization, respectively. For group II, the broken symmetry between the hBN-upper Ni(111) and hBN-lower Ni(111) bonds causes the DOS for APC to be spin-polarized. The most stable stacking arrangement DOS, in the lowest APC configuration, exhibits a spin dependence when the asymmetric arrangement is selected, as indicated in the left panel of figure \ref{fig:4}(a). This effect is caused by the pyramidal structures in the deformed BN plane. Specifically, when N atoms approach the Ni layer having the spin-up electrons of the majority spin, the N atoms are spin-polarized in the spin-down direction. The spin-polarized DOS has a larger majority spin contribution a little above the Fermi energy. This spin moment in the BN plane is, however, reversed when the N atoms approach another Ni layer with mainly the spin-down electrons. Then, the spin direction of the peak around the Fermi energy is reversed when the pyramid structure is inverted.  

The transmission calculation using Landauer-Buttiker formalism is performed for each representative of three different hybridization arrangement groups in the APC and PC states. The PWCOND module calculates the transmission probability at zero temperature and an infinitely small voltage, such that the conductance of the system can be calculated as follows

\begin{equation}
    G = \frac{I}{\delta V} = \frac{e^2}{h}T(E_F)
\end{equation}

\begin{figure}[ht]
\centering
\includegraphics[width=\linewidth]{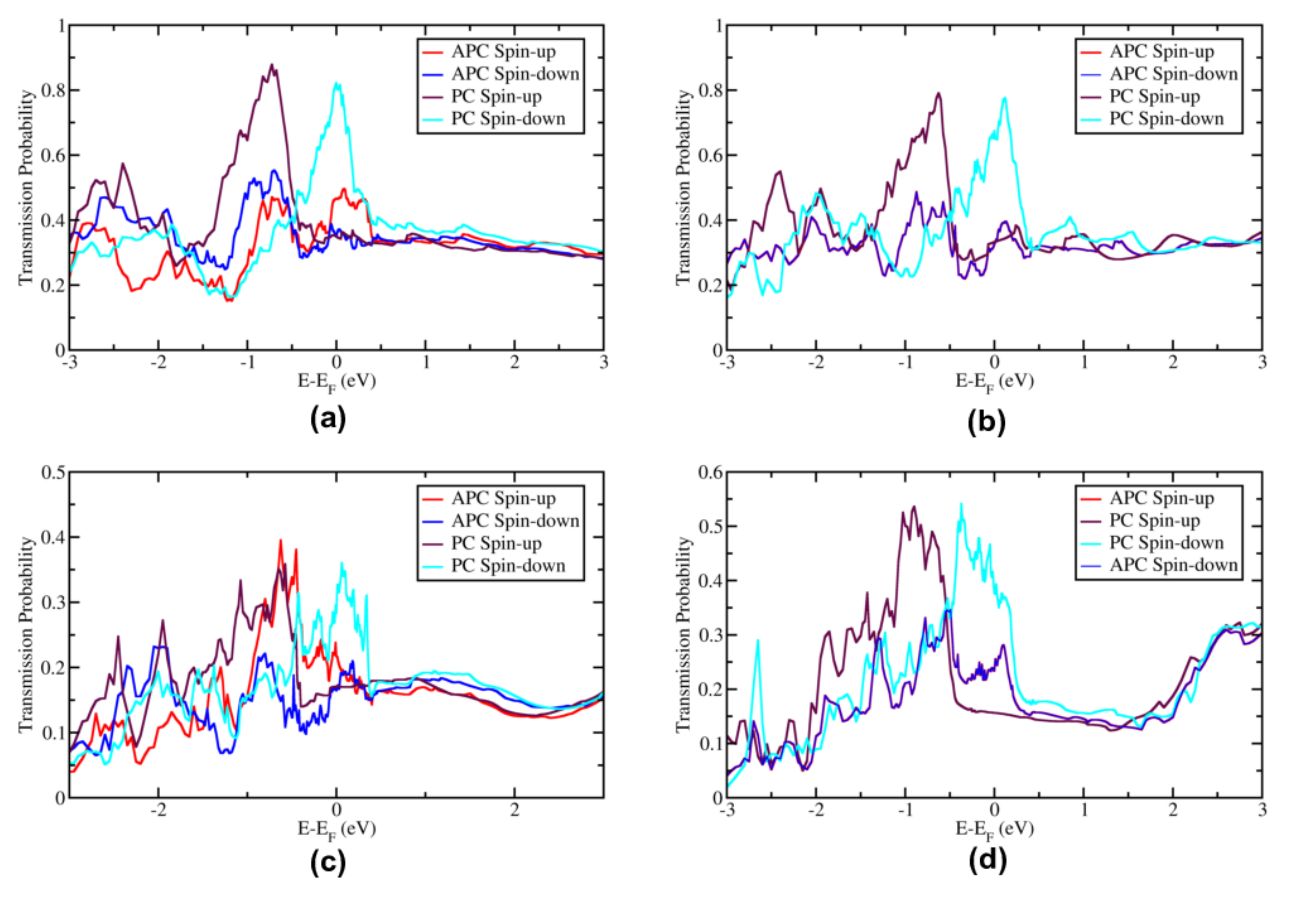}
\caption{Transmission probability of 2pd-hybridizations: (a) asymmetric and (b) symmetric, (c) 1pd-hybridization, and (d) no hybridization stacking arrangement groups in APC and PC states for spin-up and spin-down electrons.}
\label{fig:5}
\end{figure}

Figures \ref{fig:5}(a), (b), (c), and (d) show the probabilities of the scatterer, that is, group I in the asymmetric (a) and symmetric arrangements (b); group II, which is represented by the N$_{\textrm{top}}^{\textrm{fcc}}$B$_{\textrm{fcc}}^{\textrm{top}}$ stacking arrangement; and group III, represented by the N$_{\textrm{fcc}}^{\textrm{fcc}}$B$_{\textrm{top}}^{\textrm{top}}$ stacking arrangement. The transmission probability calculation indicates that only groups I and III exhibit the spin-valve effect for the APC and PC states, respectively. The absence of the spin-valve effect for group II is because of the broken symmetry in the tunneling process between Ni(111)/hBN and Ni(111) slabs. Indeed, two other stacking arrangement groups exhibit a different type of conductance. This is because the conduction path is created via bonding between Ni(111) and hBN/Ni(111) and via interlayer tunneling for groups I and III, respectively. These two different types of conductance give two different results for the transmission probability, with group I having a higher transmission probability for the spin-down electrons in the PC state, relative to that for group III. We evaluate the magnetoresistance ratio using equation \ref{eq:Julliere} in the section entitled "Computational method." The results show that the symmetric arrangement of group I produces a higher magnetoresistance than group III by a factor of 46.67\%. Although group III has a lower magnetoresistance ratio, it can change between the APC and PC states rather easily since the total energy difference between these two states is only $\approx 1.2$ meV, as can be seen in figure \ref{fig:2}. This implies that, for group III, a relatively small external magnetic field is required to change the PC state into the APC state. This functionality would allow a spin-valve device to be used as a magnetic sensor. 

The effect of symmetry on the stacking arrangement, and thus, the induced magnetic moment can be determined from the transmission probability of group I with symmetric and asymmetric stacking arrangements. The transmission probability in the APC state for either spin-up or spin-down are similar for the symmetric arrangements of the upper and lower Ni slabs with respect to the hBN layer, where the charge transfer from the Ni to N atoms does not yield an induced magnetic moment on the N atoms. Both spin configurations exhibit low transmission probabilities, corresponding to the spin blocking of the upper and lower Ni slabs.  Meanwhile, for the asymmetric arrangement, near the Fermi energy, the spin-up electrons have a higher transmission probability than the spin-down electrons. This result arises from the occurrence of an induced magnetic moment on the N atoms. This effect gives rise to another interesting function, besides spin-valve, as described in the following section.

\subsection{Tuning of spin-current by external electric field}

As explained in the previous section, the asymmetric structure of group I, the most stable state structure, exhibits a spin-dependent transmission for the lowest-energy APC state. This effect is caused by the special pyramidal structure, which applies a finite net spin moment to the N atoms of the hBN layer. Interestingly, this structure has a bi-stable local atom configuration of B$_3$N, where each of the stable pyramidal structures has a reversed form in the other structure. When the shape of the pyramidal structure is reversed, a reversal of the spin-dependent transmission probability also appears. On the other hand, the lowest asymmetric structure has charge dipoles caused by the pyramids. Thus, a structural change should occur upon the application of an electric field. In response to the electric field, the spin-dependent current changes as a result of a change in the spin-dependent transmission probability. This functionality also arises if an electromagnetic field is applied to the system, which would produce an interesting response in the spin current. The high total energy difference between the APC and PC states prevents the magnetic field component of the electromagnetic radiative field from changing the magnetic alignment configuration of the system. Therefore, only the electric field component of the electromagnetic field contributes to the change in the spin-dependent transmission probability. We may regard this correlated response as a type of cross-correlation of the electric and magnetic degrees of freedom. Thus, Ni/hBN/Ni in a 2pd-hybridization (group I) asymmetric stacking arrangement can be applied to a cross-correlation device.

\section{Conclusion}
In the present study, the spin-polarized density functional theory (DFT) calculation for Ni(111)/ hBN/Ni(111) revealed the dependencies of the magnetic properties and magnetoresistance ratio on the stacking arrangement. A total energy calculation revealed that 36 possible stacking arrangements of Ni(111)/hBN/Ni(111) can be classified into three groups: a stacking arrangement with 2pd-hybridizations (group I), with 1pd-hybridization (group II), and with no hybridization (group III). The total energies for each stacking arrangement of the two different magnetic alignment states of the upper and lower Ni(111) slabs, that is, the APC and PC states, showed that the APC state is more favorable than the PC state in a 2pd-hybridization stacking arrangement, which is in good agreement with the results of a previous experimental study \cite{Iqbal:2018-Ni-hBN-Ni_exp}. 

The charge transfer due to the hybridization of the p$_z$-orbital of N atoms with the d$_z^{2}$-orbital of the Ni atoms leads to an induced magnetic moment for the N atoms of the hBN layer. In turn, the direction of the induced magnetic moment is parallel to the Ni(111) slab with which N atoms have hybridization. Since the 36 stacking arrangements are grouped based on the configuration of the pd-hybridization, the magnetic response of the N atoms is different for each group. Group I in the PC state exhibits an induced magnetic moment on the N atoms parallel to the upper and lower Ni(111) slabs, since both hybridize with N atoms. 
On the other hand, for the APC state, the induced magnetic moment has different characteristics for the asymmetric and symmetric stacking arrangements of the Ni(111) slab. The symmetric one exhibits no net induced magnetic moment on the N atoms because the magnetic moments of the upper and lower Ni(111) slabs act in the opposite directions. 
Meanwhile, for an asymmetric stacking arrangement, a bi-stability state occurs, in which the B atoms can form a weak covalent bond with either the upper or lower Ni(111) slab. The structure in which the B atoms form a weak covalent bond with the NtopBfcc Ni(111) slab has an energy level that is $4.9$ meV lower than that of a NtopBhcp Ni(111).
Furthermore, the bi-stability state leads to a different induced magnetic response for the N atoms. The direction of the induced magnetic moment occurs in N atoms parallel to the Ni(111) slab with which the B atoms have a weak covalent bond. For group II, the N atoms only hybridize with one side of the Ni(111) slab such that the induced magnetic moment is always parallel to the Ni(111) slab with which the N atoms exhibit hybridization. For the last group, group III, since there is no hybridization, no induced magnetic moment occurs for the N atoms of hBN.

Finally, the different characteristics of hybridization in the stacking arrangement groups also affect the conductance of a spin-valve that can be observed from each transmission probability. Groups I and III exhibit a spin-valve effect in the PC and APC states, respectively. However, these two stacking arrangement groups have a different magnetoresistance ratio in which group I is higher. This difference arises from the different conduction paths on the upper and lower Ni(111) slab-hBN interface, where in group I (group III), electrons are transmitted from left to right via the interlayer covalent bonds (interlayer tunneling). However, group III can easily change between the APC and PC states according to the total energy difference relative to group I, which leads to the possibility of application of the configuration to a magnetic field sensor. On the other hand, although it is relatively difficult to change the magnetic alignment of group I from the APC to the PC state, the asymmetric stacking arrangement has a unique bi-stability state. Because of the presence of a different induced magnetic moment response on the N atoms for two different stable states, the transmission probability also changes. By applying either an electric or electromagnetic field, the polarization, as well as transmission probability, of the N and B atoms changes. Thus, it is possible to tune the spin current through the application of an external electric field. Therefore, Ni/hBN/Ni exhibits great promise for application to a cross-correlation device.

\section{Computational method}

A spin-polarized plane wave-based density functional theory calculation was performed using the Quantum ESPRESSO package \cite{qespresso:2009,qespresso:2017}. A revised Perdew-Burke-Ernzerhof (PBE) functional for a densely packed solid surface, called the PBESol functional \cite{PBEsol:2008}, within the generalized-gradient approximation, as well as ultrasoft pseudopotentials \cite{vanderbilt:1990-ultrasoft}, was used to describe the electron-ion interaction. A kinetic energy cutoff of 50 Ry was used for the wavefunctions to obtain a good convergence calculation. Since an appropriate k-point grid determined the convergence of the total energy calculation in this system, a k-point grid of 24 $\times$ 24 $\times$ 1 was used for all calculations. Furthermore, we omitted the inclusion of van der Waals interaction in the calculations, because of the presence of only a 0.02 \AA interlayer distance difference after applying this interaction. Therefore, we focused mainly on the covalent bond in the structure, and the results indicated that the main contribution to the bonding between the hBN and Ni layers originates from pd-hybridization.

All 36 Ni(111)/hBN/Ni(111) stacking arrangements were considered to determine the influence of the stacking arrangement on the magnetic properties and magnetoresistance performance. The Ni(111)/hBN/Ni(111) sandwich structure is characterized by both the stacking sequence at the interface between the hBN layer and the Ni(111) first layer and stacking sequence of the Ni(111) second and third layers, with respect to the hBN layer, with the order starting from the Ni(111) lower slab-graphene and followed by the Ni(111) upper slab-graphene, as shown in figure \ref{fig:1}.(a). For simplification, a configuration number was assigned to each stacking type, as shown in figure \ref{fig:1}. 

The ballistic transmission probability calculations were performed using Landauer-Buttiker formalism \cite{Buttiker:1985,Landauer:1970}. The left lead, right lead, and scatterer region were considered in the model calculation, as shown in figure \ref{fig:method}. Ni(111) was considered for the left and right leads to reduce the computational cost without neglecting any important physics. However, the use of Au(111) as a lead carrying normal input/output currents for a Ni/hBN/Ni structure would be expected in an actual device. The current flow from the left lead to the right lead through the scatterer can be expressed as follows:

\begin{figure}[tb]
\centering
\includegraphics[width=\linewidth]{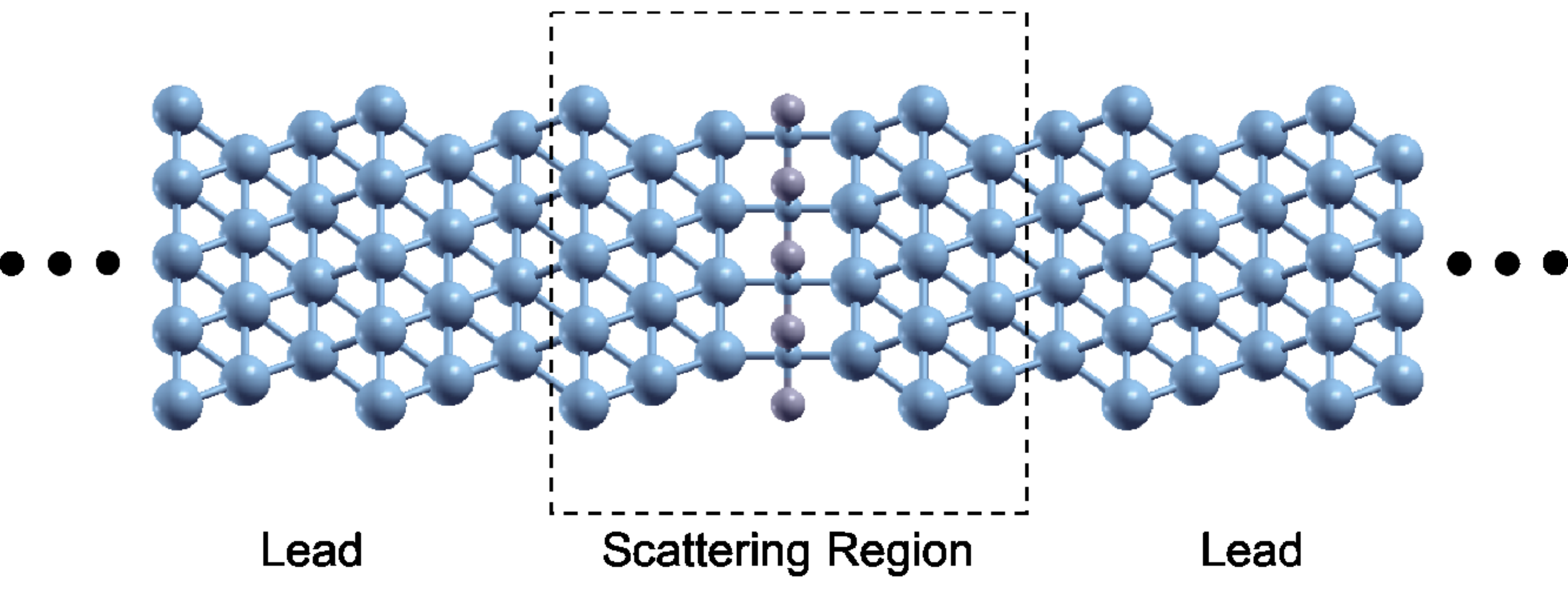}
\caption{Side view of the supercell used to represent the scattering region and lead corresponding to model calculation for transmission probability calculation.}
\label{fig:method}
\end{figure}

\begin{equation}
    I =  \frac{e}{h} \int T(E) \left[f_L(E)-f_R(E)\right] dE
\end{equation}

\noindent where $f_L(E)$ ($f_R(E)$) are right-(left-)moving electrons injected from the left (right) lead. Meanwhile, the ballistic transmission $T$ as a function of energy $E$ is described as

\begin{equation}
    T(E) = \sum_{\textbf{k}_{\parallel}} \sum_{i,j} T_{i,j}(\textbf{k}_{\parallel},E)
\end{equation}

\noindent where $T_{i,j}(\textbf{k}_{\parallel},E)$ is the probability of electrons with energy $E$ and momentum $k$, moving from Bloch state $i$-th and transmitting to Bloch state $j$-th, and then summing over the 2D Brillouin zone and all incoming-outcoming states. In the present study, the transmission probability calculations were performed using the PWCOND \cite{PWCOND:2004} module of Quantum ESPRESSO. Finally, a magnetoresistance ratio calculation was conducted based on a Julliere model \cite{Julliere:1975} by finding the difference between the conductance in the APC and PC states, and then dividing it by the conductance in the APC state, as defined by equation \ref{eq:Julliere}

\begin{equation}
    TMR = \frac{G_P - G_{AP}}{G_{AP}}
    \label{eq:Julliere}
\end{equation}

\begin{acknowledgement}

The calculations were performed at the computer centers of Kyushu University. This work was partly supported by JSPS KAKENHI Grant No.JP26400357, JP16H00914 in Science of Atomic Layers, and JP18K03456. Y. W. and H. H. gratefully acknowledge scholarship support from the Japan International Cooperation Agency (JICA) within the "Innovative Asia" Program, ID Number D1707483 and D1805252. All authors thank Dr. E. H. Hasdeo and Dr. M. K. Agusta for their fruitful discussions. 

\end{acknowledgement}

\section*{Author contributions statement}

H.H. performed the DFT calculation and spin-charge density mapping, and found the atomic configuration (arrangement) in the lowest APC state as well as the pyramidal structures of the most stable state. Y.W. applied the transmission probability calculation and found the spin-dependent current contribution in the APC state. K.K found the functionality of the APC structure. M.A.M found the consistency for different numbers of Ni layers. H.H. and Y.W. wrote the manuscript with contributions from the co-authors. All the authors reviewed the manuscript. 

\section*{Additional information}

\textbf{Competing interests:} The authors declare no competing interests. 

\bibliography{acs_ami}

\end{document}